\documentclass[11pt]{article}

\usepackage{amsmath,amssymb}
\usepackage{graphicx}
\usepackage{epstopdf}

\usepackage{epsfig}
\newcommand{\be}{\begin{equation}}
\newcommand{\ee}{\end{equation}}
\newcommand{\bea}{\begin{eqnarray}}
\newcommand{\eea}{\end{eqnarray}}
\newcommand{\ba}{\begin{array}}
\newcommand{\ea}{\end{array}}

\newcommand{\eq}[1]{(\ref{#1})}

\def\a{\alpha}      
\def\b{\beta}

\def \a {\alpha}
\def \b {\beta}

\begin{document}

\begin{titlepage}
\begin{flushright}
\normalsize {\tt hep-th/0606136}\\ Jun 15, 2006
\end{flushright}

\vfill

\begin{center}
{\large \bf
Jet Quenching Parameter in Medium with Chemical Potential 
from AdS/CFT}
\vfill
{
Feng-Li Lin\footnote{\tt linfengli@phy.ntnu.edu.tw}
}
and
{
Toshihiro Matsuo\footnote{\tt tmatsuo@home.phy.ntnu.edu.tw}
}
\vfill
{\it
Department of Physics,
National Taiwan Normal University,\\
Taipei City, 116, Taiwan
}

\end{center}

\vfill
\begin{abstract}
We calculate the jet quenching parameter in medium with chemical potential from AdS/CFT correspondence. Our result is summarized in a plot. Moreover, we extract the explicit form of the jet quenching parameter of medium with small chemical potential for phases of dual SYM corresponding to large and small black holes. For the former phase, the jet quenching is increased as the charge density increases, however, for the latter it is the opposite though the background is thermodynamically unstable.
\end{abstract}
\vfill

\end{titlepage}

\setcounter{footnote}{0}

\section{Introduction}
  
  The recent experiments in Relativistic Heavy Ion Collider (RHIC) produce the strongly interacting `quark gluon plasma', which can be described by hydrodynamical expansion of radial and elliptical flow  \cite{Kolb:2003dz}. To study the RHIC phenomenology requires the theoretical tool to calculate the dynamical process such as the diffusion constants, viscosity or jet quenching parameter in this strongly interacting plasma. Since these physical quantities characterizes the dynamical process, it is then hard to calculate them in lattice QCD which is formulated in Euclidean phase and more suitable in describing the thermal equilibrium process. On the other hand, the AdS/CFT duality \cite{Maldacena:1997re, Gubser:1998bc, Witten:1998zw} provides an avenue by using gravity/string theory to calculate the physical quantities of the strongly coupled phase of the super-Yang-Mills (SYM) theory. Even though the SYM theory and QCD are quite different at zero temperature, however, with the nonzero temperature both theory describe the similar hot non-abelian plasma's hydrodynamics except that the SYM plasma is in the adjoint representation and the QCD has only adjoint gluons and fundamental quarks. 
 In fact, in the past few years, a lot of works have been done to calculate the hydrodynamical quantities,
and one of the most important progress is the discovery of the universal bound of the viscosity to entropy density ratio \cite{Policastro:2001yc, Kovtun:2003wp, Buchel:2003tz, Kovtun:2004de, Buchel:2004qq}, which is also checked in the case of no-zero chemical potential in \cite{Mas:2006dy, Son:2006em, Maeda:2006by,Saremi:2006ep}. The check from the direct gauge theory calculation is recently done in \cite{Chen:2006ig}.
  
   The early attempt in utilizing holography to discuss the jet quenching was done by \cite{Sin:2004yx}. Recently, the jet quenching parameter has been calculated in \cite{Liu:2006ug, Herzog:2006gh, Gubser:2006bz, Casalderrey-Solana:2006rq, Herzog:2006se,Caceres, Friess:2006aw} from the gravity dual in the AdS-Schwarzschild background, and is followed up by \cite{Buchel:2006bv,Vazquez-Poritz:2006ba} for the deformed gravity background dual to less supersymmetric gauge theories. The key observation in \cite{Liu:2006ug} is that the authors define the jet quenching parameter $\hat{q}$ non-perturbatively through the relation 
\be
\langle W^A({\cal C}) \rangle = \exp \left( -{1\over 4} \hat{q} L^- L^2 \right)
\label{defjetq}
\ee
where the contour ${\cal C}$ describes the quark-anti-quark($q\bar{q}$) pairs separated by small spacelike extension of $L$ moving along the light cone of large length $L^-$. 
This relation originally arises as a dipole approximation valid for small $L$ 
used in jet quenching calculation \cite{Wiedemann:2000za,Kovner:2001vi,Zakharov:1997uu}.
It enables one to use the method of AdS/CFT duality found in \cite{Maldacena:1998im,Rey:1998ik} to calculate the expectation value of Wilson loop. On the other hand, in  \cite{Herzog:2006gh, Gubser:2006bz, Casalderrey-Solana:2006rq, Herzog:2006se, Caceres, Friess:2006aw}  the energy loss of the moving quark due to the drag forces exerted by the plasma are considered and one can extract the jet quenching parameter from the friction coefficient of the drag force. Both results agree on the dependence of the temperature and 't Hooft coupling but not on the overall coefficient.

   However, in the above calculations they ignored the effect of non-zero chemical potential except in \cite{Herzog:2006se,Caceres} in which the drag force was calculated with R-chagre chemical potential. When the $q\bar{q}$ pair is created in the vacuum they will hadronize by creating more $q\bar{q}$ pairs due to the quark confinement as they move apart, and then come out as the jets. However, in the quark-gluon plasma produced in RHIC, the escaped quark is surrounded by high density quarks fluid liberated from the nucleons of the heavy ions. Then some of the jets will be quenched by the surrounding medium and the jet quenching parameter measure the probability of the jet quenching. In such setting, the baryon density of the quark-gluon plasma is relevant when calculating the jet quenching parameter. Though there is no definitely conserved baryon number symmetry in SYM theory as in QCD, the R-symmetry plays the similar role. In this short note, we will use the AdS/CFT duality and follow the method of \cite{Liu:2006ug} to calculate the jet quenching parameter in a medium with nonzero chemical potential which is conjugated to the R-charge density. 
   
   We hope our results will help to improve the comparison between the theoretical results and experimental data.  We will first describe the gravity setting for such a calculation and then calculate the thermal expectation value of the Wilson loop to extract the quenching parameter by evaluating the on-shell Nambu-Goto action of string extending into the bulk but with endpoints fixed on the AdS boundary. 
   Our result can be summarized in Figure 1 which shows the ratio between the jet quenching parameters with and without chemical potential at the same temperature. 
   The explicit form of the jet quenching parameter in some limiting cases is also derived.

\section{Calculation of the Jet quenching parameter }
 The $SU(N)$ SYM theory with non-zero chemical potential background is dual
to the bulk 5-dimensional gauged supergravity in the asymptotically AdS
R-charged black hole background, some subtleties about the phase structures of the theory is recently discussed in \cite{Yamada:2006rx}. The background metric of the single R-charge black hole is
\cite{Son:2006em,Behrndt:1998jd,Chamblin:1999tk,cvetic99}
\footnote{
This is obtained by dimensional reduction on $S^5$ from the 10-dimensional metric of spinning near-extremal 3-brane background of IIB supergravity, see \cite{Behrndt:1998jd,Chamblin:1999tk,cvetic99} for details. Therefore our treatment here is equivalent to the full 10-dimensional setting. The same reduction was adopted in \cite{Son:2006em} to discuss the viscosity of quark-gluon plasma. 
}%
\bea
ds^2_5 &=& - H^{-2/3}{(\pi T_0 R)^2 \over u}\,f \, dt^2
+  H^{1/3}{(\pi T_0 R)^2 \over u}\, \left( dx^2 + dy^2 + dz^2\right)
+ H^{1/3}{R^2 \over 4 f u^2} du^2\,,
\\\nonumber
&=&-H^{-2/3}{(\pi T_0 R)^2 \over u}(H+f)dx^+dx^-+{1\over2} H^{-2/3}
{(\pi T_0 R)^2 \over u}(H-f)\left( (dx^+)^2+(dx^-)^2\right)
\\
&&+ H^{1/3}{(\pi T_0 R)^2 \over u}\, \left( dx^2 + dy^2\right)+ H^{1/3}
{R^2 \over 4 f u^2} du^2\,,
\\\nonumber
&:=&G_{MN}dx^M dx^N
\label{metric}
\eea
where $R$ is the AdS radius, the functions $f$, $H$ and the parameter $T_0$ are
\begin{equation}
f(u) = H(u) - u^2 (1+\kappa)\,,
\;\;\;\; H= 1 + \kappa u \,, \;\;\;\;
\kappa \equiv {q\over r_+^2}\,, \;\;\;\; T_0 = r_+/\pi R^2\,.
\label{def1}
\end{equation}
Here $q$ is related to the physical charge of the black hole, and
$r_+$ is largest root satisfying $h(r)=0$ where $h(r):=1-{q\over
r^2}+{r_0^4 \over r^4}$ is the harmonic function 
in the Schwarzschild
coordinate $r$ which is related to the above coordinate $u$ by
$u=r^2_+/r^2$. Thus, the black hole horizon is at $u=1$ and the AdS
boundary at $u=0$.

Note that the above
solution is obtained from the AdS charged non-extremal spherical
black hole by blowing up the sphere of its horizon. The non-extremal
parameter $r_0$ is related to the black hole mass parameter $m$ of
the spherical black hole by $r_0^4=mR^2$.

Moreover, there also configurations of scalar fields and gauge fields
but their explicit form will be omitted. For simplicity, we will consider the
case with $\kappa_1=\kappa$ but $\kappa_2=\kappa_3=0$. In this case, the physical
parameters in SYM theory can be related to the parameters in the
supergravity background. The temperature of SYM is the Hawking temperature
\be
T_\kappa={2+\kappa \over 2\sqrt{1+\kappa}} T_0.\label{th1}
\ee
The density of physical charges is
\be
\rho={\pi N^2 T_0^3\over 8} \sqrt{2\kappa(1+\kappa)}
\ee
where $N$ is rank of the gauge group
and the chemical potential conjugated to $\rho$ is
\be
\mu=\pi T_0 \sqrt{2\kappa \over 1+\kappa}.\label{mudef}
\ee
Moreover, the parameter can be written as 
$\kappa=8\pi^2 \rho^2/s^2$ where $s$ is the
entropy density, and the background is thermodynamically stable only
if
\be
\kappa<2.
\ee

   Now we would like to evaluate the on-shell action of a string
worldsheet ending on the AdS boundary and extending into the bulk,
\be
S={1\over 2\pi \alpha'} \int d\sigma d\tau \sqrt{\det g_{\alpha \beta}}
\ee
where $g_{\alpha \beta}=G_{MN}\partial_{\a}x^M\partial_{\b}x^N$ is
the pull-back metric. Then the on-shell worldsheet action is related
to the thermal expectation value of the Wilson-loop for SYM in the
fundamental representation \cite{Maldacena:1998im, Rey:1998ik} by
\be
\langle W^F({\cal C})\rangle =\exp[-S({\cal C})] \label{wsmetric}
\ee
where the contour ${\cal C}$ on the AdS boundary enclosed the worldsheet 
surface of the string.

To mimic the $q\bar{q}$ pairs moving in the hot quark-gluon plasma along
the lightcone of length $L^-$, the contour ${\cal C}$ should be
lightlike with large extension of size $L^-$ in the $x^-$-direction
and small extension of size $L (\ll L^-)$ in a transverse direction,
i.e. $x$-direction. For such a contour, it is convenient to
parametrizing the worldsheet by target space coordinates $(x^-=\tau,
x^+=const, x=\sigma, y=const, u=u(\sigma))$ by assuming the
worldsheet is time translationally invariant. 
Given this, the
worldsheet action (\ref{wsmetric}) takes the form
\bea
S
&=& {L^-\sqrt{1+\kappa}\over \sqrt{2} \pi \alpha'}  (\pi T_0 R)^2
{1\over 2\pi T_0} \int^{L/2}_0 d\sigma {u'\over \sqrt{H^{1/3}u f}}
\sqrt{1+(2\pi T_0)^2 {u f \over u'^{\;2}}} .
\label{S12}
\eea
This can be compared with the
action for the trivial configuration given by two disconnected
worldsheet running from $u=0$ to $u=1$, namely,
\be
S_0= {L^-\sqrt{1+\kappa}\over \sqrt{2} \pi \alpha'}  (\pi T_0 R)^2
{1\over 2\pi T_0} \int^1_0 {du\over \sqrt{H^{1/3}u f}}\label{S00}
\ee
which represents the infinite bare mass of the $q\bar{q}$ pair and
should be subtracted off from the action (\ref{S12}) of the on-shell
configuration.

  From (\ref{S12}) the equation of motion is
\be
H^{1/3}\left( 1+{1\over (2\pi T_0)^2} {u'^{\;2}\over u f}
\right)=\frac{1}{E^2}
\label{eom1}
\ee
where $E^2$ is the constant of motion. Using (\ref{eom1}) to eliminate
$u'$ in (\ref{S12}) and assuming that $E^2\ll1$ (low energy), then we arrive the
following subtracted action
\be
S_I:=S|_{on-shell} - S_0 \approx {L^-\sqrt{1+\kappa}\over 2\sqrt{2}
\pi \alpha'}  (\pi T_0 R)^2{E^2\over 2\pi T_0}
\int^1_0  {du\over \sqrt{H^{-1/3}u f}} \label{Si0}.
\ee
Note that sign $\approx$ reminds already using the condition
$E^2\ll1$.

  To evaluate the subtracted action $S_I$, we need to solve the
equation of motion (\ref{eom1}). For the string extending from AdS
boundary at $(u=0, \sigma=-L/2)$, bending over at the horizon at
$(u=1, \sigma=0)$ and then coming back to AdS boundary at $(u=0,
\sigma=-L/2)$, the equation (\ref{eom1}) can be put into the following
form 
\be
{L\over 2}= {1\over 2\pi T_0} 
\int_0^1 {du \over \sqrt{(E^{-2}H^{-1/3}-1)u f}}
\approx {E\over 2\pi T_0} \int_0^1 {du \over \sqrt{H^{-1/3} u f }}.
\label{eomsol}
\ee
Again, we have used $E^2 \ll1$ to arrive the second expression, which
relates $E$ to $T_0$, $L$ and $\kappa$. From \eq{eomsol} the condition 
$E^2\ll 1$ holds if $T_0L\ll 1$ which is reasonable since $L$ is extremely small for jet quenching.  

Using (\ref{eomsol}) to eliminate $E^2$ in (\ref{Si0}), we arrive
\be
2 S_I\approx {\pi^2 \over 4 \sqrt{2}}\sqrt{\lambda}T_0^3 L^-L^2
Q(\kappa)
\label{sif}
\ee
where $\lambda=R^4/\alpha'^{\;2}$ is the 't Hooft's coupling of SYM
theory, and
\be
Q(\kappa):=2\sqrt{1+\kappa}
\left[\int_0^1 \frac{du}{\sqrt{H^{-1/3}uf}}\right]^{-1} ,
\label{Q}
\ee
which is a monotonically increasing function.

Following \cite{Liu:2006ug} we take
\bea
\hat{q}_{YM} \equiv -\frac{4}{L^- L^2} \ln \langle W^A({\cal{C}})\rangle 
\eea 
as a non-perturbative definition of the jet quenching parameter.
Therefore we find 
\bea
\hat{q}_{YM}(\kappa,T_\kappa,\lambda)
&=&\frac{\pi^2}{\sqrt{2}} \sqrt{\lambda}T_0^3 Q(\kappa)
\nonumber \\
&=&\frac{\pi^2}{\sqrt{2}} \sqrt{\lambda}
\left(\frac{2\sqrt{1+\kappa}}{2+\kappa}\right)^3T_\kappa^3 Q(\kappa)
\label{formula}
\eea
where we have converted $T_0$ to the physical temperature $T_\kappa$, 
and this is our main result in this note.

It is easy to see that the subtracted action (\ref{sif}) reduces to the
result in \cite{Liu:2006ug} for the zero r-charge case, i.e., $\kappa=0$.
Therefore, the jet quenching parameter
$\hat{q}_{YM}(\kappa,T_\kappa,\lambda)$ with
chemical potential characterized by $\kappa$ is related to the one without chemical potential $\hat{q}_{YM}^{(0)}(T_0,\lambda)$ by
\be
\hat{q}_{YM}(\kappa,T_\kappa,\lambda)
={Q(\kappa)\over Q(\kappa=0)}\;\hat{q}_{YM}^{(0)}(T_0,\lambda) ,
\ee
where
\bea
\hat{q}_{YM}^{(0)}(T_0,\lambda)=\frac{\pi^2}{\sqrt{2}} \sqrt{\lambda}T_0^3 Q(0) .
\eea
It is interesting to compare these two parameter at the same physical temperature, 
namely $T_\kappa=T_0$.
The ratio should be multiplied by a factor and one can see 
\bea
\frac{\hat{q}_{YM}(\kappa,T_\kappa=T_0)}{\hat{q}_{YM}^{(0)}(T_0)}
=\left(\frac{2\sqrt{1+\kappa}}{2+\kappa}\right)^3\frac{Q(\kappa)}{Q(0)} .
\eea 
We plot the ratio as a function of $\kappa$ in Figure 1. 

The plot shows the background charges increase the jet quenching parameter 
for small $\kappa$ whereas decrease for large $\kappa$. Recall that $\kappa:=q/r_+^2$ where $r_+$ is the horizon size, so for fixed charge $q$, the small $\kappa$ corresponds to large black hole and vice versa. The results then suggest that the jet quenching of the dual SYM's corresponding to large and small black holes have opposite charge dependences. Especially for the small black hole's SYM dual, the jet quenching is smaller for larger amount of charge, this is at odd with the naive expectation.  Note also that the dividing point is around $\kappa=2$, it is interesting to see if this is related to the thermodynamically instability or not.
\begin{figure}[ht]
\center{\epsfig{figure=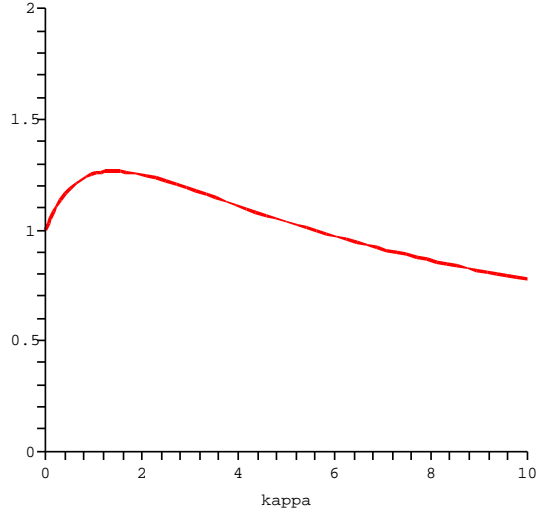,angle=0,width=8cm}}
   \caption{The y-axis is the ratio between jet quenching parameters with and without
chemical potential $\hat{q}_{YM}(\kappa,T_\kappa)/\hat{q}_{YM}^{(0)}(T_0)$ at $T_\kappa=T_0$, 
and the x-axis is the parameter $\kappa$ characterizing the charge-to-entropy density ratio.}
   \label{fig}
\end{figure}

  To extract the explicit form of $\kappa$ dependence in
$Q(\kappa)$,  we can consider either $\kappa\ll 1$ limit or
$\kappa\gg1$ limit, though as we mentioned before the background is thermodynamically unstable for $\kappa >2$.
We investigate it for the theoretical interest.

  Note that for $\kappa \ll 1$, from \eq{th1} and \eq{mudef} we have
\be
T_0 \approx T_{\kappa} + {\cal O}(\kappa^2), \qquad \kappa \approx {1\over 2\pi^2}\left( {\mu \over T_{\kappa}} \right)^2 \ll 1.
\ee
So, this is the case with small chemical potential.  Similarly, for $\kappa \gg 1$, 
 \be
 T_0\approx {2\over \sqrt{\kappa}} T_{\kappa}, \qquad \kappa\approx \left({2\pi \sqrt{2} T_{\kappa} \over \mu} \right)^2 \gg 1 \label{kgg1}
\ee
this also implies small chemical potential. So, in both $\kappa \gg 1$ and $\kappa \ll 1$ limits, the chemical potential is small compared with the Hawking temperature. 

In the limit for $\kappa$ small, there is no difference in $T_\kappa$ and $T_0$ up to the  linear order in $\kappa$, and one can easily find from (\ref{Q})
\bea
Q(\kappa) \simeq Q(0)(1+c_1 \kappa +O(\kappa^2)) .
\eea
where $c_1$ is given by 
%
\bea
c_1&=&
\frac{3}{4}-\frac{10\pi^2}{3\Gamma(1/4)^4}
\nonumber \\
&\simeq& 0.5596 .\label{c1c}
\eea
Thus the ratio of the quenching parameters is given as
\bea
\frac{\hat{q}_{YM}(\kappa,T_\kappa=T_0)}{\hat{q}_{YM}^{(0)}(T_0)}
=1+c_1 \kappa +O(\kappa^2) .
\eea
This shows that the effect of the chemical potential introduced by the background charge increase the quenching parameter.

For large $\kappa$, $Q(\kappa)$ behaves as 
\bea
Q(\kappa) \simeq d_1 \kappa^{7/6}
\eea
where $d_1 = \pi^{-3/2} \Gamma(2/3)\Gamma(5/6) \sim 0.7275$.
Thus we find 
\bea
\hat{q}_{YM}(\kappa,T_\kappa) \simeq \frac{8\pi^2 d_1}{\sqrt{2}} \sqrt{\lambda}
T_\kappa^3 \kappa^{-1/3} .
\eea
The temperature and chemical dependence is peculiar when expressing $\kappa$ in terms of $T_{\kappa}$ and $\mu$ by \eq{kgg1}. 

\section{Conclusion} 
In this note we calculated the jet quenching parameter in a medium with nonzero chemical potential. We find that the phases of the gauge theory dual to large and small black holes have opposite charge dependence, especially, the result for the  latter is out of expectation as the jet quenching decreases as the charge increases. Moreover, the temperature and 
chemical potential dependence 
of the jet quenching for the small black hole case is also peculiar.  

%
\begin{table}[htdp]
\begin{center}
\begin{tabular}[t]{|c|c|c|c|} 
\hline     & T=300 & T=400  & T=500        \\  
\hline $\kappa=0$&3.17 & 7.51 &14.7      \\ 
\hline $\kappa=1$&3.97 & 9.42 &18.4      \\ 
\hline $\kappa=10$&2.46 & 5.83 &11.4   \\ 
\hline
\end{tabular}
\end{center}
\caption{The jet quenching parameter 
$\hat{q}_{YM}$ GeV$^2$/fm for various values of $\kappa$ with $\lambda=6 \pi$.}
\label{default}
\end{table}%

We like to remind the reader that the black hole background is thermodynamically unstable for $\kappa >2$ so that our results in Figure 1 and Table 1 for the regime of $\kappa>2$ should be taken with caution. Despite that we think that it is still interesting to have a look of jet-quenching in this regime since the quark-gluon plasma in RHIC is also in a meta-stable state, and could be mimicked by the meta-stable phase of $\kappa>2$. The lifetime of this meta-stable phase and the comparison with the RHIC data would be an interesting dynamical issue for further investigation. 

It is tempting to compare our results with the experimental data and may need to involve subtle experimental data analysis which could be beyond our reach. Moreover, the R-charge in SYM is only qualitatively mimicking the baryon number in QCD because these two theory have different field contents under these symmetries. Instead
we give some numbers from our formula (\ref{formula}) in Table 1
which might be comparable to the experimental data. However, we like 
to mention
that the constant $c_1$ obtained in \eq{c1c} is positive so that the charge density helps to quench the jet. This is consistent with what has been expected from the estimate in \cite{Liu:2006ug} where 
without taking the chemical potential into account,
the authors found that the theoretical jet quenching parameter is smaller than the estimated experimental data 
We will leave the detailed comparison to the future when the more experimental data for jet quenching parameter appear. Finally, we would like to mention that our result is qualitatively similar to the one in \cite{Herzog:2006se} from the drag force calculation in the sense that the jet quenching reaches maximum at $\kappa={\cal O}(1)$, but quantitatively different in the detailed dependence of $\kappa$. 

\bigskip

{\bf Note added in proof:} After we submitted our paper, the papers \cite{sfetsos06,mas06} appeared, which discuss the same problem as in ours but in the context of 10-dimensional spinning 3-brane. However, the results in those papers agree with ours qualitatively. 

\section*{Acknowledgements}
We would like to thank Juinn-Wei Chen for very inspiring discussions on this project, and also thank Tetsuo Hatsuda, Chung-Wen Kao, Akitsugu Miwa, Eiji Nakano, Dan Tomino, Wen-Yu Wen for discussions on related issues, and Chris Herzog for helpful comments. 
FLL also likes to thank the hospitality of SIAS, Fundan Univ. and CMS of Zhejiang Univ. where part of the work is completed.
This work is supported by Taiwan's NSC grant 94-2112-M-003-014.


\end{document}